\documentclass[12pt]{article}

\usepackage{epsfig}
\usepackage{amsfonts}
\usepackage{amsmath}
\usepackage{amssymb}
\usepackage{bm}

\newcommand{\ep}{\epsilon}

\newcommand{\al}{\alpha}
\newcommand{\bt}{\beta}

\newcommand{\Fmu}{F_{\mu\nu}}

\newcommand{\slashPT}{\slash\hspace{-0.6em}P\slash\hspace{-0.5em}T}
\newcommand{\slashPTsub}{\slash\hspace{-0.45em}P\slash\hspace{-0.4em}T}

\newcommand{\Nb}{\bar N}
\newcommand{\Fp}{F_\pi}

\newcommand{\tb}{\bar \theta}

\newcommand{\mpi}{m_{\pi}}
\newcommand{\MQCD}{M_{\mathrm{QCD}}}

\newcommand{\Or}{\mathcal O}

\newcommand{\dslash}[1]{#1 \llap{/\kern-0.5pt}}
\newcommand{\Dslash}[1]{#1 \llap{/\kern+1.2pt}}
\newcommand{\DDslash}[1]{#1 \llap{/\kern+2.3pt}}
\newcommand{\dslashh}[1]{#1 \llap{/\kern+1pt}}

\newcommand{\boldtau}{\mbox{\boldmath $\tau$}}
\newcommand{\boldpi}{\mbox{\boldmath $\pi$}}

\begin{document}

\begin{titlepage}

\vspace{2.0cm}

\begin{center}
{\Large\bf Deuteron Magnetic Quadrupole Moment \\ \vspace{0.2cm}
From Chiral Effective Field Theory}

\vspace{1.7cm}

{\large \bf C.-P. Liu$^1$,  
J. de Vries$^2$, E. Mereghetti$^{3}$,\\
\vspace{0.2cm}
R. G. E. Timmermans$^2$, and U. van Kolck$^{4,5}$}

\vspace{0.7cm}
{\large 
$^1$ 
{\it Department of Physics, National Dong Hwa University, 
\\
Shoufeng, Hualien 97401, Taiwan}}

\vspace{0.25cm}
{\large 
$^2$ 
{\it KVI, Theory Group, University of Groningen,
\\
9747 AA Groningen, The Netherlands}}

\vspace{0.25cm}
{\large 
$^3$ 
\it Ernest Orlando Lawrence Berkeley National Laboratory,\\
University of California, Berkeley, CA 94720, USA}

\vspace{0.25cm}
{\large 
$^4$ 
\it Institut de Physique Nucl\'{e}aire, 
Universit\'e Paris-Sud, IN2P3/CNRS,\\
F-91406 Orsay Cedex, France
}

\vspace{0.25cm}
{\large 
$^5$ 
\it Department of Physics, University of Arizona,
\\
Tucson, AZ 85721, USA}

\end{center}

\vspace{1.5cm}

\begin{abstract}
We calculate the magnetic quadrupole moment (MQM) of the deuteron at
leading order in the systematic expansion provided by chiral effective field
theory. We take into account parity ($P$) and time-reversal ($T$)
violation which, at the quark-gluon level, results from the QCD vacuum
angle and 
dimension-six operators that originate from physics beyond
the Standard Model. We show that the deuteron MQM can be expressed in
terms of five low-energy constants that appear in the  $P$- and $T$-violating
nuclear potential and electromagnetic current, four of which also contribute
to the electric dipole moments of light nuclei. We conclude that the deuteron MQM
has an enhanced sensitivity to the QCD vacuum angle and that its measurement would be complementary to the proposed
measurements of light-nuclear EDMs.
\end{abstract}

\vfill
\end{titlepage}

Permanent electric dipole moments (EDMs) of particles, nuclei, atoms, 
and molecules
are powerful probes for physics ``beyond'' the Standard Model (SM) of particle
physics. EDMs violate both parity ($P$) and time-reversal ($T$) invariance 
(or, if we invoke the $C\!PT$ theorem of relativistic quantum field theory, 
$C\!P$ invariance).
EDMs could be low-energy manifestations of some source of $P$ and $T$ violation 
($\slashPT$) 
that originates at an energy scale comparable to 
or even higher
than that accessed by the LHC. At the current experimental accuracy, 
electroweak quark
mixing can 
be ignored, and the only Standard Model (SM) source that can 
impact EDMs is the QCD
vacuum angle, $\tb$ \cite{dim4}. 
A nonzero measurement of a hadronic EDM
could be due to either beyond-the-SM $\slashPT$ sources or
a finite value of $\tb$, even though the current upper
bound on the neutron EDM already limits this
angle to a very small value. 
Among the possible beyond-the-SM $\slashPT$ sources,
those represented
by operators of effective dimension six 
\cite{dim6}
are expected to dominate:
the quark EDM (qEDM), 
the quark and gluon chromo-electric dipole moments (qCEDM and gCEDM, 
respectively), 
and two independent four-quark (FQ) interactions.
From a theoretical point of view, it is clearly an important priority 
to disentangle 
the dimension-four SM $\tb$-term and the non-SM sources \cite{Vri11b}.

The deuteron is an excellent candidate for a sensitive EDM search 
in a storage ring \cite{Far04}.
In a recent paper \cite{Vri12} we addressed the EDMs of light nuclei, 
including the deuteron,
within chiral effective field theory (EFT). 
The major advantage of 
such an EFT approach is its
direct link to QCD in that it exploits the different chiral properties
of the fundamental $\slashPT$ sources. 
Moreover, the power-counting scheme allows 
for a 
controlled framework such
that the theoretical uncertainties can be estimated and the results 
can be improved systematically. 
We showed that the EDMs of light nuclei can be expressed 
in terms of essentially 
six $\slashPT$ parameters, or low-energy constants (LECs). 
These LECs can in principle
be calculated from the underlying $\slashPT$ sources by solving QCD 
at low energies,
in particular by lattice simulations. 
Lacking that, the size of the LECs can be estimated
by naive dimensional analysis. 
We concluded that the EDMs of various light nuclei can give
crucial complementary information about the fundamental 
$\slashPT$ sources. 

Since it is a spin-1 particle, the deuteron has one other static $\slashPT$ 
electromagnetic moment, the magnetic quadrupole moment (MQM). 
In this paper, we address the deuteron
MQM in the same framework that was used in Ref.\,\cite{Vri12}. 
It was shown in
Ref.\,\cite{Vri11b} that, in addition to LECs that contribute 
to the EDMs of light nuclei,
the deuteron MQM depends also on $\slashPT$ pion-nucleon-photon interactions. 
Moreover, it was argued that only for the $\tb$ term is the deuteron MQM 
expected
to be larger than the deuteron EDM (in appropriate units). 
For the beyond-the-SM sources,
the MQM is expected to be of similar size or somewhat smaller than the EDM. 
This indicates
that a measurement of the deuteron MQM, if possible, could play a central role 
in separating the various $\slashPT$ sources.

The conclusions of Ref.\,\cite{Vri11b} were based on a chiral EFT 
in which the one-pion
exchange nucleon-nucleon ($N\!N$) force is treated 
in perturbation theory \cite{KSW}. 
This approach,
expected to be valid at low energies, allows one to give analytical 
results for the deuteron
EDM and MQM, which, moreover, can be extended to sub-leading orders. 
On the other
hand, in $N\!N$ scattering it was found that the results do not converge 
in certain partial
waves for momenta somewhat lower than expected \cite{Fleming99}. 
It is therefore important to check
the results of Ref.\,\cite{Vri11b} 
in the framework of  a chiral EFT that treats pions
nonperturbatively \cite{nonpert}. 
For the deuteron EDM it was found that the two EFTs
gave similar results \cite{Vri12}. 
In this article we investigate this for the deuteron MQM as
well and we address the question, to what extent a possible 
measurement of the MQM
could be of help to separate the different $\slashPT$ sources. 
We also compare our results
to previous studies of the deuteron MQM \cite{Khr00,Liu04}. 
In particular, Ref.\,\cite{Liu04}
used traditional meson-exchange $N\!N$ models and a general $\slashPT$ $N\! N$ 
interaction \cite{Liu04,Liu06}. 
We use the codes of Ref.\,$\cite{Liu04}$, but adapt and
extend the framework (the $\slashPT$ $N\! N$ potentials and currents) 
to chiral EFT with nonperturbative pions. We follow the hybrid approach of Ref. \cite{Vri12}
in which nonperturbative pion exchange
is embedded into
the $P$- and $T$-conserving ($PT$) wave functions of modern, ``realistic''
potentials \cite{Wir95, Sto94}.
In principle our framework can be applied to 
the calculation of the MQM of other light nuclei as well.

We start our discussion of the deuteron MQM from the 
effective hadronic Lagrangian involving
the low-energy degrees of freedom:
nucleon ($N$), photon ($A_\mu$), and pion ($\boldpi$).
The $PT$ part, 
which originates from the quark kinetic (color-
and electromagnetically gauged) and mass terms in QCD, is well-known
\cite{ChPTreview, Vri12}.  
At leading order (LO) 
it consists of the standard pion-nucleon axial-vector coupling, 
$g_{A}=1.27$, and the pion-nucleon-photon interaction obtained from gauging the 
$g_A$ term. The pion-photon interactions stem from the pion charge. 
At next-to-leading order (NLO),
the photon couples to the nucleon via 
the covariant derivative in the nucleon kinetic term
and via the isoscalar
and isovector magnetic moments, respectively
$\kappa_0=-0.12$ and  $\kappa_1 = 3.7$. 
The $\slashPT$
effective Lagrangian results from the most general  
$\slashPT$ Lagrangian at the quark-gluon level. 
It was derived and discussed in detail in Refs.\,\cite{Mer10,Dim6}.
The key observation is that the different fundamental 
$\slashPT$ sources transform
differently under chiral symmetry and therefore generate 
different hadronic interactions.
Given enough independent observables it will be possible to disentangle 
them \cite{Vri11b}.

We present here only the terms that are relevant for the LO MQM
calculation, which depends on five $\slashPT$ interactions. The Lagrangian
\begin{eqnarray}\label{LPTV}
{\cal L}_{\slashPTsub} & = & -\frac{1}{F_\pi} \bar{N}
 \left(\bar{g}_{0}\,\boldtau\cdot\boldpi+\bar{g}_{1} \pi_3 \right) N
  +\frac{\bar c_\pi}{\Fp} \ep^{\mu \nu \al \bt}  v_\al  \, 
   \Nb S_\bt \boldtau\cdot\boldpi N  \,
   \Fmu \nonumber \\ 
&& + \bar C_{1} \bar N\!N \partial_{\mu} (\bar N S^{\mu} N)  
  + \bar C_{2} \bar N \boldtau N \cdot \mathcal D_{\mu} (\bar N S^{\mu} \boldtau N) \ ,
\label{LslashT}
\end{eqnarray}
contains isoscalar ($\bar{g}_{0}$) and isovector ($\bar{g}_{1}$) nonderivative
pion-nucleon couplings, an isoscalar pion-nucleon-photon coupling 
($\bar c_\pi$),
and two short-range $\slashPT$ $N\!N$ interactions ($\bar C_{1}$, $\bar C_{2}$).
Here $F_{\pi}=185$ MeV is the pion decay constant,
$\boldtau$ are the Pauli matrices in isospin space,
$v_\mu$ ($S_\mu$) is the nucleon velocity (spin),
$\ep^{\mu \nu \al \bt}$ (with $\ep^{0123} = 1$)
the completely antisymmetric tensor,
$F_{\mu\nu}=\partial_\mu A_\nu -\partial_\nu A_\mu$
the photon field strength,
and $(\mathcal D_\mu)_{ab} = \delta_{ab} \partial_\mu + e \epsilon_{3ab} A_\mu + \Or(\boldpi \cdot \partial_\mu \boldpi/\Fp^2)$ 
a chiral covariant derivative. 
Except for $\bar c_\pi$, these interactions
play a role in the calculation of nuclear EDMs \cite{Vri11b,Vri12,Vri11a}
as well.
The scaling of the LECs in terms of 
the pion mass ($m_\pi$), 
the characteristic QCD scale ($M_{\rm QCD} \sim 2\pi F_\pi$), 
and the 
scale of $\slashPT$ physics
beyond the SM is given in Ref.\,\cite{Vri12} 
for $\bar{g}_0$, $\bar{g}_1$,
$\bar{C}_1$, and $\bar{C}_2$. 
The pion-nucleon-photon coupling $\bar c_\pi$  has the same scaling 
as the short-range contribution to the isoscalar nucleon EDM, 
$\bar d_0$ \cite{Vri11b}.

The calculation of the deuteron MQM can be divided into two contributions. 
The first contribution comes from an insertion of the $\slashPT$ electromagnetic
two-body current $\vec{J}_{\slashPTsub}$. The current has to be two-body since the
constituent nucleons, being spin-$1/2$ particles, do not possess a MQM. 
The second contribution comes from the electromagnetic current $\vec{J}_{PT}$ 
upon perturbing the wave function of the nucleus with the $\slashPT$ potential 
$V_{\slashPTsub}$, 
such that the wave function obtains a $\slashPT$ component. This current can be
one- or two-body. 
The required $\slashPT$ potential $V_{\slashPTsub}$ and current 
$\vec{J}_{\slashPTsub}$ can be calculated from Eq.\,(\ref{LPTV}). 

To first order in the $\slashPT$ sources, the deuteron MQM
is thus a sum of two reduced matrix elements,
\begin{equation}
\mathcal M_d=\frac{1}{\sqrt{30}}\left(
\langle \Psi_{d}||\widetilde{\bm{M}}||\Psi_{d}\rangle
+2\,\langle \Psi_{d}||\bm{M}||\widetilde{\Psi}_{d}\rangle \right)\,.
\end{equation}
The deuteron ground state $|\Psi_{d}\rangle$ and its parity
admixture $|\widetilde{\Psi}_{d}\rangle$ are the solutions of 
homogeneous and inhomogeneous Schr\"odinger equations,
\begin{eqnarray}
(E-H_{PT})|\Psi_{d}\rangle & =& 0 \ , 
\\
(E-H_{PT})|\widetilde{\Psi}_{d}\rangle & = &
V_{\slashPTsub}|\Psi_{d}\rangle \ ,
\end{eqnarray}
respectively, where $H_{PT}$ is the $PT$ Hamiltonian. 
The MQM operators $\bm{M}$ and $\widetilde{\bm{M}}$
are obtained from the corresponding currents
$\vec{J}_{PT}$ and $\vec{J}_{\slashPTsub}$, respectively. The Cartesian component  
along the $z$ direction, $M_{33}$, which is 
proportional to the spherical harmonic $Y_2^0$, takes the form  
\begin{equation}\label{Mzz}
M_{33} = 2\,\int d^3x\,x_3\,(\vec{x}\times\vec{J}(\vec{x}))_3\,,
\end{equation}
where $\vec{x}$ is the position where the current density is probed.
Given the current in momentum space,
\begin{equation}
\vec{J}(\vec{q}) = \int d^3x\,e^{-i\,\vec{q}\cdot\vec{x}}\,\vec{J}(\vec{x})\,,
\end{equation}
$M_{33}$ can also be derived as 
\begin{equation}
M_{33} = -2\,\lim_{\vec{q}\rightarrow0}\left(\nabla_{q_3} \nabla_{q_1} J_2(\vec{q}) 
- \nabla_{q_3} \nabla_{q_2} J_1(\vec{q})\right)\,.
\end{equation} 

\begin{figure}[t]
\centering
\includegraphics[scale = 0.6]{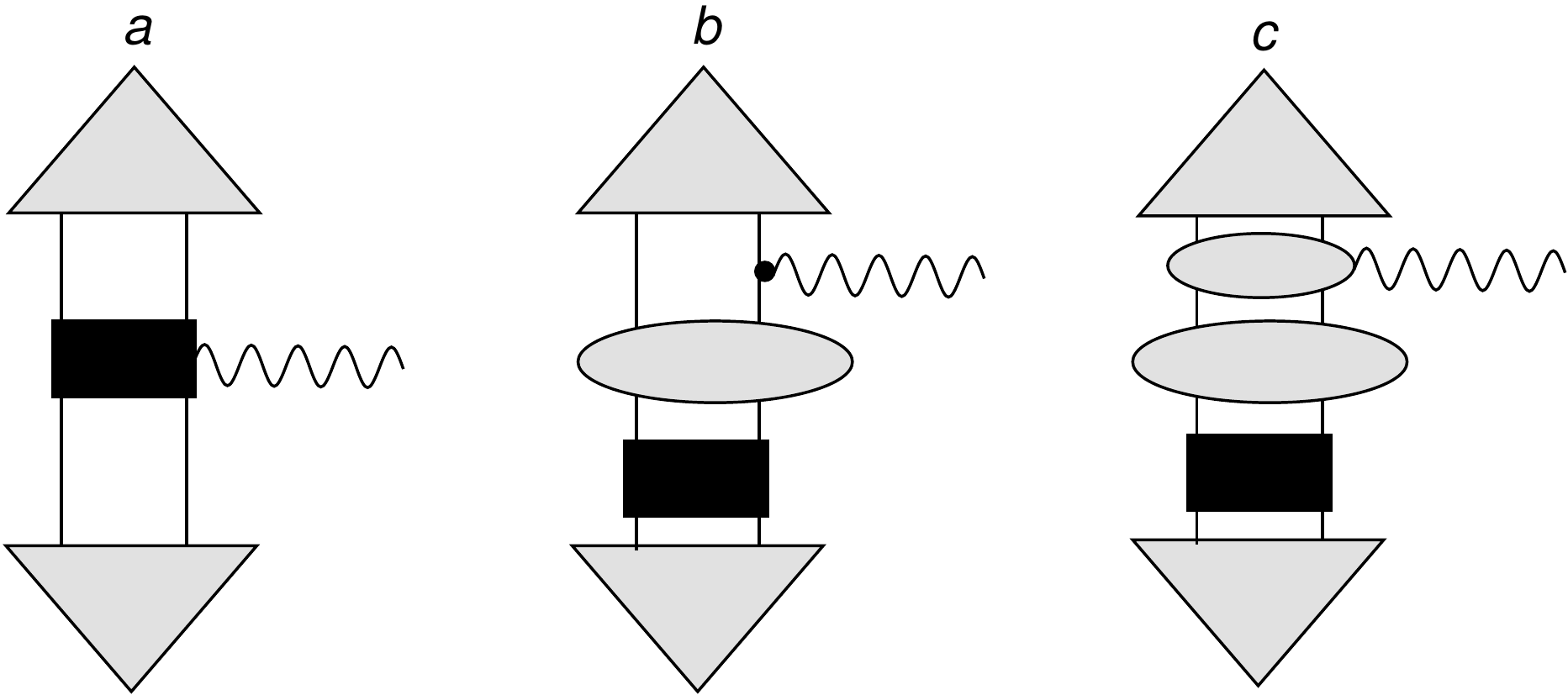}
\caption{The three general classes of diagrams contributing to the deuteron MQM
described in the text. Solid and wavy lines represent nucleons and photons. The
large triangle denotes the 
$PT$ wave function and the oval iterations of the $PT$ potential. 
The nucleon-photon
vertex is the nucleon magnetic moment or the convection current,
and the photon attached to the oval is
the $PT$ two-body current.
The black squares without (with) an attached photon
represent the $\slashPT$ potential 
(two-body current). 
}
\label{classesMQM}
\end{figure}

The three classes of contributions to the deuteron MQM described above are 
shown in Fig.\,\ref{classesMQM}. 
In order to decide which diagrams 
give the main contribution to the MQM we apply 
the power counting rules outlined in Ref. \cite{Vri12}. These rules provide an expansion in $Q/\MQCD$, where 
$Q$ is the generic momentum in the process, in this case of the
order of the deuteron binding momentum. We use $Q\sim\mpi\sim\Fp$, as standard in $\chi$PT. 
For each class we take the $PT$ vertices from the
interactions described above and 
the $\slashPT$ 
LO vertices from Eq.\,(\ref{LPTV}). 
The iteration of the LO $PT$ potential is not suppressed,
and is necessary among nucleons in reducible intermediate
states, as indicated in diagrams (b) and (c) of Fig.\,\ref{classesMQM}.
Such iteration among nucleons before and after any $\slashPT$
insertion builds up the $PT$ wave function,
represented in  Fig.\,\ref{classesMQM} by the triangles. 
In the following power counting we omit this overall factor.
The scaling in terms of the LECs in Eq.\,(\ref{LPTV}) of diagram (a) is
then 
\begin{equation}\label{Ma}
D_a = 
{\cal O} \left(e\,\frac{\bar g_{0}}{F_\pi^2} \frac{Q^2}{M_{\textrm{QCD}}} \right)
+{\cal O} \left(\frac{\bar c_\pi Q^2}{F_\pi^2}  \frac{Q^2}{M_{\textrm{QCD}}}\right)
+{\cal O} \left(e\,\bar C_{2} \Fp^2 \frac{Q^2}{M_{\textrm{QCD}}}\right),
\end{equation} 
while diagrams (b) and (c) scale as
\begin{equation}\label{Mb}
D_{b,c} = 
{\cal O} \left(e\,\frac{\bar g_{0,1}}{F_\pi^2} \frac{Q^2}{M_{\textrm{QCD}}} \right)
+{\cal O} \left(e\,\bar C_{1,2} \Fp^2 \frac{Q^2}{M_{\textrm{QCD}}} \right).
\end{equation}

Which contribution dominates
depends on the fundamental $\slashPT$ source,
the dimension-four $\tb$-term and the dimension-six terms: 
qEDM, qCEDM, gCEDM, and FQ.

\begin{itemize}
\item For the $\tb$ term, only the isoscalar coupling $\bar g_0$ 
plays a role at LO.
In order to generate $\bar g_1$, the $\tb$ term requires an insertion 
of the quark mass difference, which causes a relative suppression of 
$\bar g_1$ relative to 
$\bar g_0$ by a factor $\varepsilon \mpi^2/\MQCD^2$ \cite{Mer10}, 
where $\varepsilon =(m_d-m_u)/(m_u+m_d)$.
The other couplings, $\bar c_{\pi}$ and $\bar C_{1,2}$, also 
contribute at sub-leading orders. 
Since $\bar g_0$ enters in principle through the three classes of diagrams with
similar factors, all these classes
are equally important.

\item For the qCEDM, we need both pion-nucleon interactions, since there
is no relative suppression of $\bar g_{1}$.
Again, all three classes
are, a priori, equally important.

\item For the qEDM, the purely hadronic interactions are suppressed 
by factors of the fine-structure constant, and
only $\bar c_\pi$ is important.
Thus only diagram (a) matters. 

\item For the chiral-invariant ($\chi$I) $\slashPT$ sources (gCEDM and FQ), 
the pion-nucleon interactions, which break chiral symmetry, are suppressed 
by a factor $\mpi^2/\MQCD^2$ compared to the $\slashPT$ 
short-range $N\!N$ interactions, 
which conserve chiral symmetry. The latter
are therefore 
as important as the pion-nucleon interactions.
For the $\chi$I sources again all three diagrams could be important. 
\end{itemize}

We now turn to the calculation of the various
ingredients in these diagrams,
starting with the currents.
For the nucleons we use
incoming momenta $\vec{p}_{1}=\vec{P}/2+\vec{p}$ and 
$\vec{p}_{2}=\vec{P}/2-\vec{p}$,
and outgoing momenta $\vec{p}_{1}^{\,'}=\vec{P}^{\,'}/2+\vec{p}\,'$ and 
$\vec{p}_{2}^{\,'}=\vec{P}^{\,'}/2-\vec{p}\,'$.
The photon momentum $\vec{q}=\vec{P}-\vec{P}^{\,'}$ is outgoing. 
For convenience
we introduce $\vec{k}=\vec{p}-\vec{p}\,'$. 
The spin (isospin) 
of nucleon $i$ is denoted by $\vec\sigma^{(i)}/2$ ($\boldtau^{(i)}/2$).
\begin{figure}[t]
\centering
\includegraphics[scale = 0.6]{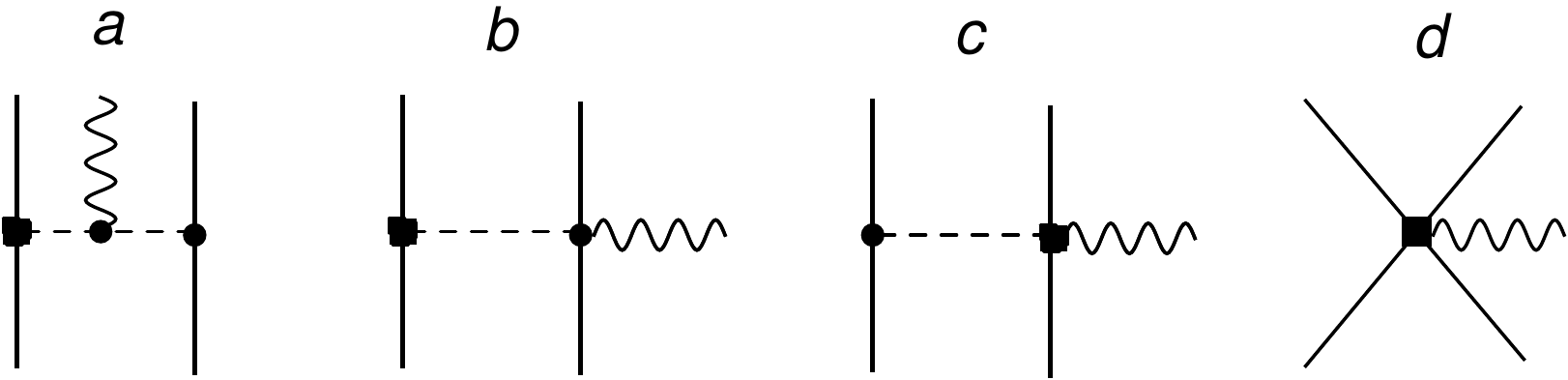}
\caption{Diagrams contributing to the $\slashPT$ two-nucleon  current.
Solid, dashed, and wavy lines represent nucleons, pions, and photons.
A square marks a $\slashPT$ interaction  
and the other
vertices $PT$ interactions.
Only one topology per diagram is shown.}
\label{MQMcurrentsTodd}
\end{figure}

\begin{figure}[t]
\centering
\includegraphics[scale = 0.6]{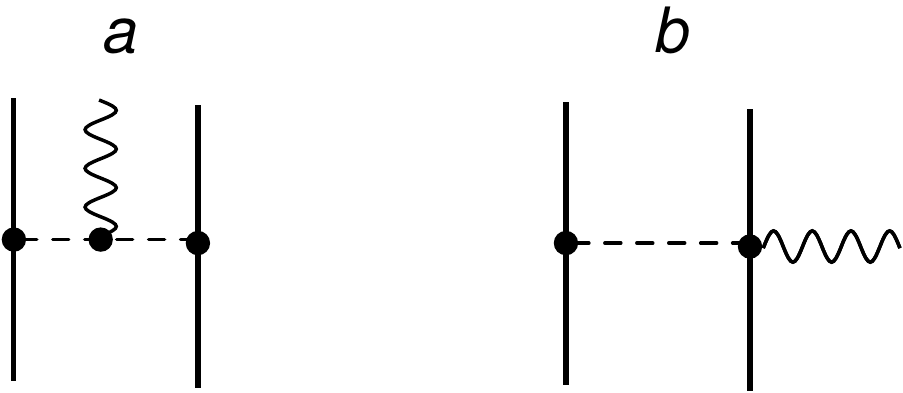}
\caption{Diagrams contributing to the $PT$ two-nucleon current.
}
\label{MQMcurrentsTeven}
\end{figure}

The relevant $\slashPT$ two-body currents 
that appear in diagram class (a) of Fig.\,\ref{classesMQM} 
are shown in Fig.\,\ref{MQMcurrentsTodd}.
Since the deuteron wave function is isoscalar, the $\slashPT$ currents 
need to be isoscalar as well in order to contribute to the MQM. 
Only the current in diagram (c), which stems
from qEDM, meets this requirement, 
and we find
\begin{equation}\label{slashPTcurrent}
\vec{J}_{\slashPTsub
}(\vec q, \vec k) =
-\frac{g_A \bar c_\pi}{\Fp^2} \boldtau^{(1)}\cdot\boldtau^{(2)} 
\bigg[\vec \sigma^{(1)}\times \vec{q} 
\,\frac{\vec \sigma^{(2)}\cdot (\vec k -\vec q/2)}{(\vec k -\vec q/2)^2+\mpi^2} 
-\vec \sigma^{(2)}\times \vec{q}
\,\frac{\vec \sigma^{(1)}\cdot (\vec k +\vec q/2)}{(\vec k +\vec q/2)^2+\mpi^2}
\bigg].
\end{equation}
In diagram classes (b) and (c) of Fig.\,\ref{classesMQM} 
the photon interacts instead with a $PT$ current.
The $PT$ one-body current in diagram (b) 
is either 
the nucleon magnetic moment or the convection current coming from the 
nucleon kinetic energy,
\begin{equation}
\vec{J}_{PT}(\vec q, \vec p_i)
=\frac{e}{4m_N} 
\left\{\left[1+\kappa_0+\left(1+\kappa_1\right)\tau^{(i)}_3\right]
i\vec \sigma^{(i)}\times \vec q
+ \left(1 + \tau_3^{(i)}\right)\,
\left(2\vec{p}_i
-\vec{q}\right) \right\},
\label{1BPTcurrent}
\end{equation}
where $i$ is the index 
of the nucleon that interacts 
with the photon. 
In diagram (c) we require the $PT$ two-body currents
depicted in Fig.\,\ref{MQMcurrentsTeven}: 
\begin{eqnarray}
\vec{J}_{PT}(\vec q, \vec k) &=& 
i \frac{e g_A^2}{\Fp^2} \left(\boldtau^{(1)}\times \boldtau^{(2)}\right)_3
\left\{-2
\vec{k}\,
\frac{\vec \sigma^{(1)}\cdot(\vec k+\vec{q}/2)}{(\vec k +\vec q/2)^2+\mpi^2} \,
\frac{\vec \sigma^{(2)}\cdot(\vec k -\vec q/2)}{(\vec k -\vec q/2)^2+\mpi^2}  
\right.
\nonumber\\
&& \left.
+\vec \sigma^{(1)}\,
\frac{\vec \sigma^{(2)}\cdot(\vec k-\vec q/2)}{(\vec k -\vec q/2)^2+\mpi^2}+
\vec \sigma^{(2)}\,
\frac{\vec \sigma^{(1)}\cdot(\vec k+\vec q/2)}{(\vec k +\vec q/2)^2+\mpi^2} 
\right\} . 
\label{2BPTcurrent-ab}
\end{eqnarray}

With the methods outlined in 
Refs. \cite{Koelling:2011mt, Vri12},
these currents can straightforwardly be Fourier 
transformed to coordinate space,
where we denote by $\vec x^{(i)}$ the position of nucleon $i$ 
and by $\vec r = \vec x^{(1)}-\vec x^{(2)}$ the relative position.
We also introduce the Yukawa function
\begin{equation}
U(r)=\frac{e^{-m_{\pi}r}}{4\pi r}\, .
\end{equation}
The MQM operators are, from Eq. (\ref{slashPTcurrent}),
\begin{eqnarray}
\widetilde{M}_{33} &=& 
2\,\frac{g_A \bar{c}_{\pi}}{\Fp^2} \boldtau^{(1)}\cdot\boldtau^{(2)}
\left[\left(3\,\sigma_3^{(1)}x_3^{(1)}-\vec{\sigma}^{(1)}\cdot\vec{x}^{(1)}\right)
\vec{\sigma}^{(2)}  
\right.
\nonumber \\ 
& & \left.
- \left(3\,\sigma_3^{(2)}x_3^{(2)}-\vec{\sigma}^{(2)}\cdot\vec{x}^{(2)}\right) 
\vec{\sigma}^{(1)}\,\right]
\cdot\vec{\nabla}_rU(r)\,,
\end{eqnarray} 
and from Eqs. \eqref{1BPTcurrent} and \eqref{2BPTcurrent-ab},
\begin{eqnarray}
M_{33} &=& \frac{e}{2m_N} \left\{
\left[1+\kappa_0+(1+\kappa_1)\tau^{(i)}_3\right] 
\left(3\, \sigma^{(i)}_3 x^{(i)}_3-\vec{\sigma}^{\,(i)}\cdot\vec{x}^{\,(i)}\right) 
\right.
\nonumber \\
& & \left.
+ 2 \left(1+\tau^{(i)}_3\right) 
\vec{x}^{(i)}_3\,(\vec{x}^{(i)}\times\vec{p}^{(i)})_3\right\}
\nonumber \\
&& -\frac{eg_A^2}{2\Fp^2} \left(\boldtau^{(1)}\times \boldtau^{(2)}\right)_3 
\left\{
\left[ \sigma^{(1)}_3\,\left(\vec{\sigma}^{(2)} \times \vec{r} \right)_3 
+ \sigma^{(2)}_3\,\left(\vec{\sigma}^{(1)} \times \vec{r} \right)_3\right] 
\right.
\nonumber \\
&& \left.
+4
\Big[ x^{(1)}_3 \left(\vec{x}^{(1)}\times \vec{\sigma}^{(1)} \right)_3 
\,\vec{\sigma}^{(2)} 
+ x^{(2)}_3 \left(\vec{x}^{(2)} \times \vec{\sigma}^{(2)}\right)_3 
\,\vec{\sigma}^{(1)} 
\Big] \cdot \vec{\nabla}_r 
\right\} U(r)\,,
\end{eqnarray} 
where $\vec{p}^{(i)}=-i \vec{\nabla}_{x^{(i)}}$.

The last ingredients we need are the 
deuteron wave function $\Psi_d$ and its parity admixture $\tilde \Psi_d$. They 
have been calculated in Ref.\,\cite{Liu04} using modern high-quality 
phenomenological $PT$ potentials and a $\slashPT$ potential dominated by 
pion exchange and extended with heavy-meson exchange. 
In the 
EFT spirit, it would be best
to calculate $\Psi_d$ from a $PT$ Hamiltonian fully 
consistent with chiral symmetry and renormalization-group invariance, but 
an accurate fit to two-nucleon data with these properties does not yet exist.
For those sources for which the MQM is dominated by long-distance physics,
we expect no significant differences when 
using the phenomenological potentials,
and thus for $\Psi_d$ we use the results 
of Ref.\,\cite{Liu04}. For comparison
with Ref. \cite{Vri12}, we give below numbers corresponding to
the Argonne $v_{18}$ (AV18) potential \cite{Wir95}. Differences with results obtained from 
the NijmII and Reid93 potentials \cite{Sto94} are within 
a few percent, except for the cases of $\bar{C}_{1,2}$ insertion---for which 
more details will be provided later. 

The ground state of the deuteron is mainly a $^{3}S_{1}$ state with 
some ${}^3D_1$ admixture.
The matrix element of $\widetilde{\bm{M}}$ is found to be
\begin{equation}\label{cpi}
\frac{1}{\sqrt{30}}
\langle \Psi_{d}||\widetilde{\bm{M}}||\Psi_{d}\rangle = 
0.07 \frac{\Fp \bar c_\pi}{e}\; e\,\mathrm{fm}^2.
\end{equation}

The relevant matrix elements of $\bm{M}$ are obtained 
with $\tilde \Psi_d$ from 
the LO $\slashPT$ 
two-nucleon potential. The general
$\slashPT$ $N\!N$ potential was derived in Ref.\,\cite{Mae11} 
and we summarize the relevant
parts here. 
In coordinate space the potential is given by
\begin{eqnarray}
V_{\slashPTsub}(\vec r) 
&=&  -  \frac{\bar g_0 g_A}{F^2_{\pi}} 
\boldtau^{\, (1)} \cdot \boldtau^{\, (2)} 
\left( \vec{\sigma}^{\, (1) }-\vec{\sigma}^{\,  (2) }\right) \cdot 
\left(\vec{\nabla}_r\, U(r)\right) 
\nonumber \\ 
& &
-\frac{\bar g_1 g_A}{2 F^2_{\pi}}
\left[ \left(\tau_3^{(1)} + \tau_3^{(2)} \right)
\left( \vec\sigma^{(1)}  -  \vec\sigma^{(2)}\right) 
+\left(\tau_3^{(1)} - \tau_3^{(2)} \right) 
\left( \vec\sigma^{(1)} +  \vec\sigma^{(2)}\right) \right] \cdot 
\left(\vec  \nabla_r U(r) \right)
\nonumber \\
& & + \frac{1}{2} 
\left[\bar C_{1}+\bar C_{2} \boldtau^{\, (1)} \cdot \boldtau^{\, (2)}\right]
\left( \vec{\sigma}^{\, (1) }-\vec{\sigma}^{\,  (2) }\right) 
\cdot \left(\vec \nabla_r\, \delta^{(3)}(\vec r\,)\right).
\end{eqnarray}
In this expression, at LO $\bar g_0$ originates
from $\tb$ term, qCEDM, and $\chi$I sources;  $\bar g_1$ from qCEDM and 
$\chi$I sources;
and $\bar C_i$ from $\chi$I sources only.

If the parity admixture comes from a $\bar{g}_{1}$ pion exchange, 
the deuteron wave function acquires a $^{3}P_{1}$ component. 
In order to get back to the ground state, 
the deuteron can couple to the photon via the isovector one-body or the 
two-body currents in Eqs. \eqref{1BPTcurrent} and \eqref{2BPTcurrent-ab},
respectively.
The result is
\begin{equation}\label{g1}
\frac{2}{\sqrt{30}}
\langle \Psi_{d}||\bm{M}||\widetilde{\Psi}_{d}({}^3 P_1)\rangle = 
-\bigg[0.031(1+\kappa_1)+ 0.003+0.008 \bigg]
\frac{\bar{g}_{1}}{\Fp} \; e\,\mathrm{fm}^2,
\end{equation}
where the first and second terms come from the isovector magnetic moment 
due to the spin and convection current, respectively; 
and the third term
from the $PT$ two-body current. 

\begin{figure}[t]
\centering
\includegraphics[scale = 0.5]{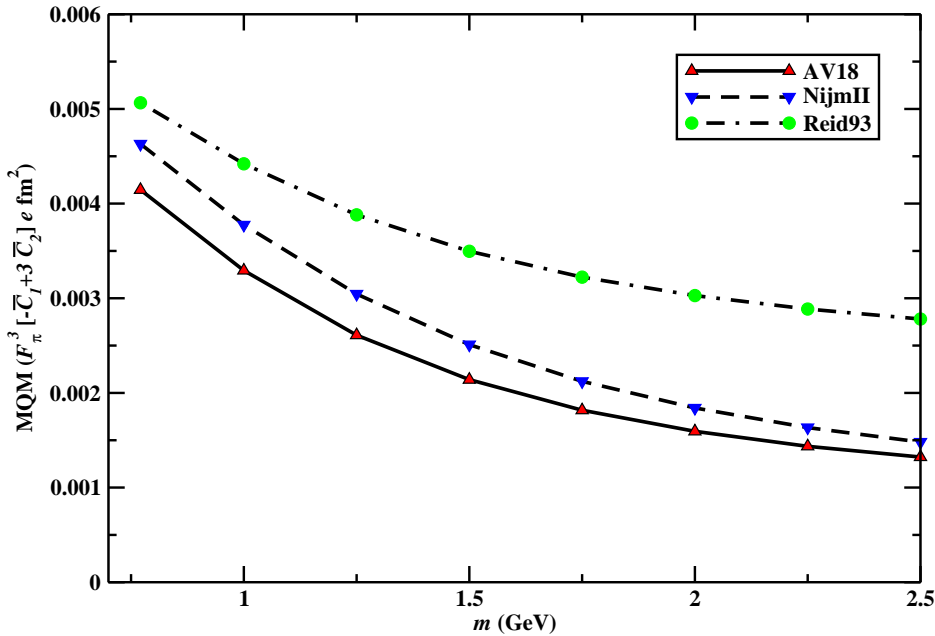}
\caption{Deuteron MQM due to the $F_\pi^3(-\bar C_1 +3\bar C_2)$ 
short-range $\slashPT$ interaction as function
of a regulating mass $m$, for various $PT$ Hamiltonians.
}
\label{C1C2_plot}
\end{figure}

After a $\bar{g}_{0}$ pion exchange or an
insertion of $\bar{C}_{1,2}$
the deuteron wave function obtains a $^{1}P_{1}$ component instead.
In this case, the deuteron needs to couple 
to the isoscalar nucleon magnetic moment or an isoscalar two-body current, 
which is not present at LO. The result is
\begin{equation}\label{g0C0}
\frac{2}{\sqrt{30}}
\langle \Psi_{d}||\bm{M}||\widetilde{\Psi}_{d}({}^1 P_1)\rangle = 
-\bigg[0.044  \,\frac{\bar{g}_{0}}{\Fp} 
+ 0.0013 \Fp^3\left(\bar C_1 -3\bar C_2\right)\bigg]
(1+\kappa_0)\; e\,\mathrm{fm}^2.
\end{equation}
For
the contact interaction with 
\begin{equation}
\bar C_0 \equiv\bar C_1 -3\bar C_2 
\end{equation}
we apply a strategy followed in Refs. \cite{CPonP,Vri12}:
it is simulated in our calculations by a fictitious
heavy-meson (of mass $m$) exchange, since
\begin{eqnarray}
\frac{m^2 \bar{C_0}}{4\pi r}e^{-m r}  &\to&
\bar{C_0}\delta^{(3)}(\vec{r}\,)
\end{eqnarray}
as $m$ goes to infinity. 
As shown in Fig.\,\ref{C1C2_plot}, when $m$ reaches $2.5\,\mathrm{GeV}$, 
the results converge at about $\lesssim 10\%$ level, 
so we report the above numbers at this scale.
While in this figure one sees good consistency between the AV18 and NijmII 
results, 
the Reid93 result is off by a factor of $2$. 
The main reason is that the Reid93 potential generates 
a deuteron $S$ state whose short-distance wave function is 
enhanced, 
leading to more sensitivity to
$\bar{C_0}$.
The large discrepancy between different potentials suggests 
that for $\chi$I sources,
for which $\bar C_0$ contributes to the MQM at leading order, 
a fully consistent calculation of $\Psi_d$ within EFT
is necessary
if this part of the matrix element needs to be known
better than within a factor of $2$.

The dependence of the deuteron MQM on $\bar g_{0,1}$, $\bar C_0$, 
and $\bar c_\pi$ was studied in Ref. \cite{Vri11b} 
in a framework where pion exchange is treated perturbatively. 
At LO in that framework, the coefficients in front of 
$\bar g_0(1+\kappa_0)/\Fp$ and $\bar g_1(1+\kappa_1)/\Fp$ were found to be, 
respectively, $-0.146$ and $-0.049$, in agreement with the results in Ref. \cite{Khr00} where a zero-range approximation for the $N\!N$ interaction was assumed.
Considering the large intrinsic uncertainty ($\sim 30\%$) 
in the perturbative-pion calculation, 
the perturbative-pion 
$\bar g_1$ coefficient is in reasonable agreement with Eq. \eqref{g1}. 
A similar agreement was found for the deuteron EDM \cite{Vri12}. 
On the other hand, the perturbative-pion $\bar g_0$ coefficient 
is three times larger than Eq. \eqref{g0C0}, 
suggesting that the effects of additional pion exchanges, 
neglected in the LO perturbative calculation, are
larger in the $^{1}P_{1}$ channel than in the $^{3}P_{1}$ channel. 
We have verified that, if the tensor force is ignored 
and the same strong force is assumed for both the $^1P_1$ and $^3P_1$ channels,
the ratio of the strong parts
of the MQM matrix elements 
due to the isoscalar and isovector one-body currents 
becomes 3,
which is consistent with Refs. \cite{Khr00,Vri11b}. 
Preliminary results of
an NLO calculation 
in the perturbative-pion framework
indicate that, indeed, NLO corrections influence the $\bar g_0$ coefficient 
by a larger amount than the $\bar g_1$ coefficient \cite{NLO}.

In the framework of perturbative pions, the $PT$ 
convection and two-body currents in diagram (b) and (c) of 
Fig. \ref{classesMQM} enter at NLO and are expected to be smaller 
than the contribution from the isovector magnetic moment. 
This agrees with the numerical results in Eq. \eqref{g1} 
where the convection and two-body currents only 
enter at the $\sim 5\%$ level. 
These currents would have been small ($\sim 30\%$) even if the 
isovector magnetic moment had been more natural,
that is, if $1+\kappa_1$ were $\simeq 1$.

The result for $\bar c_\pi$ in Eq. \eqref{cpi} is somewhat smaller 
than expected from the power counting estimate in Eq. \eqref{Ma}, 
$\mathcal M_d(\mathrm{qEDM}) \sim 0.2 \Fp \bar c_\pi \,\mathrm{fm}^2$,
and is more in line with the expectation of the perturbative-pion calculation,
$\mathcal O( \gamma \bar c_{\pi}/ M_{NN} M_{\textrm{QCD}} ) 
\sim 0.07 \Fp \bar c_\pi \,\mathrm{fm}^2$, 
where $M_{N\!N} = 4\pi \Fp^2/g_A^2 m_N$ is the 
characteristic
scale where pions become nonperturbative \cite{KSW} and $\gamma$ the deuteron binding momentum.
A more detailed comparison with the perturbative-pion calculation is 
complicated by the appearance of a short-range current, 
which is needed for renormalization purposes. 
Neglecting the counterterm and using for the renormalization scale 
$\mu = M_{N\!N}$, the perturbative result becomes 
$0.08 \Fp \bar c_\pi \,\mathrm{fm}^2$, 
in good agreement with Eq. \eqref{cpi}.
A comparison between the contributions
from the short-range $\slashPT$ $N\!N$ interactions
in the perturbative and nonperturbative calculations 
is not useful because
the LEC $\bar C_0$ includes different physics
and thus has different scalings in the two EFTs. 

The deuteron MQM was previously calculated in Refs. \cite{Khr00, Liu04}, in which the deuteron MQM was assumed to be dominated by $\slashPT$ one-pion exchange (OPE). Since these calculations did not use the chiral properties of the fundamental 
$\slashPT$ sources, the $\slashPT$ pion-nucleon interactions were assumed 
to be all of the same size. Our analysis shows that these assumptions only hold in case of a qCEDM. For $\bar g_0$ and $\bar g_1$ OPE we confirm the results in Ref. \cite{Liu04}, but our $\bar g_0$ result is $3$ times smaller than the result in Ref. \cite{Khr00} due the discrepancy discussed above.  

We are now in the position to discuss the results for the various 
$\slashPT$ sources. It was noted in Ref. \cite{Vri11b} 
that the observation of a deuteron MQM, 
together with the nucleon and deuteron EDMs, could provide
important clues to separate the various $\slashPT$ sources. 
In particular, it was concluded that only for chiral-symmetry-breaking, 
but isoscalar, sources like the QCD $\bar \theta$ term
is the deuteron MQM, in appropriate units, substantially larger
than the deuteron EDM. For chiral- and isospin-breaking sources, 
like the qCEDM, the deuteron MQM and EDM are expected to be of 
the same size, and a measurement of both would fix the couplings 
$\bar g_0$ and $\bar g_1$, allowing a prediction of other 
$\slashPT$ observables, like the $^3$He EDM.
For the $\chi$I sources and the qEDM, in the perturbative-pion approach 
the MQM depends on one- and two-body LECs that do not contribute to the EDM. 
For these sources the MQM was found to be of the same size, 
or slightly smaller, than the EDM, but an observation of the MQM 
would not give us predictive power.
These conclusions, based on the perturbative-pion power counting, 
are confirmed here by the nonperturbative results.

For the QCD $\bar \theta$ term, the deuteron EDM is dominated by 
the isoscalar nucleon EDM,  $d_d \sim 2 \bar d_0$ \cite{Vri11b, Vri12}. 
A naturalness lower bound on the isoscalar nucleon EDM is provided by the 
non-analytic terms stemming from the pion cloud \cite{EDMNLO}, 
$|\bar d_0| \gtrsim 0.01 (|\bar g_0|/\Fp)\,e\, \mathrm{fm}$. 
The deuteron MQM is dominated by the $\bar g_0$ piece in Eq. \eqref{g0C0}.
Combining the two with the deuteron mass $m_d$, we find
\begin{eqnarray}
\left|\frac{m_d \mathcal M_d}{d_d}\right|
&\simeq& 0.21 (1+\kappa_0) \left|\frac{\bar{g}_0}{\Fp \bar d_0}\right|
\,e\, \mathrm{fm}
\lesssim  21 (1+\kappa_0) .
\end{eqnarray}
At LO in the perturbative-pion approach, this ratio 
is about three times larger, as discussed above. 
Nonetheless, the nonperturbative calculation confirms that for 
isoscalar chiral-breaking sources the deuteron MQM is expected
to be larger than the EDM in units of $m_d$.

In case of the qCEDM, where $\bar g_0$ and $\bar g_1$ have similar scalings, 
both the deuteron EDM and MQM are dominated by pion exchange. 
At LO, the EDM depends on $\bar g_1$ only \cite{Vri12}, 
and the MQM on the $\bar g_{0,1}$ contributions 
in Eqs. \eqref{g1} and \eqref{g0C0}. The MQM/EDM ratio becomes
\begin{eqnarray}
\frac{m_d \mathcal M_d}{d_d} &\simeq& 
1.6(1+\kappa_1) + 2.2 (1+\kappa_0)\frac{\bar g_0}{\bar g_1} +0.6,
\end{eqnarray}
which formally is $\Or(1)$. However, due to the large anomalous isovector 
magnetic moment, numerically the ratio could be $\Or(10)$. 
This means that the measurement of a large MQM/EDM ratio does not 
necessarily imply that the $\tb$ term is the dominant $\slashPT$
mechanism.

If a qEDM is the dominant $\slashPT$ source,
the MQM is given by Eq. \eqref{cpi}. 
It is solely coming from a two-body $\slashPT$ current. 
The EDM is given by the sum of the neutron and the proton EDM,
$d_d\simeq 2 \bar d_0$ \cite{Vri12}.
Combining these results gives
\begin{eqnarray}
\frac{m_d \mathcal M_d}{d_d}& \simeq &0.7 \frac{ \bar c_\pi}{\bar d_0},
\end{eqnarray}
which is $\Or(1)$ by naive dimensional analysis (NDA). 
As we observed in the previous discussion, the matrix element of the operator 
with coefficient $\bar c_{\pi}$ is smaller than the power counting estimate, 
so a more accurate conclusion is that, for $\slashPT$ from the qEDM, 
the deuteron MQM is expected to be slightly smaller than the EDM, 
in agreement with Ref. \cite{Vri11b}.

Finally, for chiral-invariant sources, the MQM is dominated by the sum 
of Eqs. \eqref{g1} and \eqref{g0C0}. 
From the NDA estimates in Ref. \cite{Vri12}, 
we find $\Fp^4 \bar C_{0}/\bar g_0 = \Or(\Fp^2/\mpi^2) \simeq 2$, 
such that the $\bar C_0$ contribution only enters at the $\sim 10\%$ level. 
The deuteron EDM at LO formally depends on $\bar g_{1}$ and 
the isoscalar nucleon EDM, but numerically the latter 
is expected to dominate,
with pion-exchange corrections at the $\sim 15\%$ level \cite{Vri12}. 
Ignoring the numerically small convection and two-body currents,
\begin{eqnarray}
\left|\frac{m_d \mathcal M_d}{d_d}\right|
&\simeq& \left[0.21 (1+\kappa_0) \left|\frac{\bar{g}_0}{\Fp \bar d_0}\right|
+ 0.15(1+\kappa_1)\left|\frac{\bar{g}_1}{\Fp \bar d_0}\right|
\right]\,e\, \mathrm{fm}.
\end{eqnarray}
By power counting we would expect the ratio to be $\Or(1)$, 
but since the deuteron is weakly bound, 
pion exchange is smaller than expected.
Using the NDA estimate
$|\bar d_0|\sim 5 (|\bar g_{0,1}|/\Fp) \, e\,\mathrm{fm}$,
we conclude that for $\chi$I sources the deuteron MQM should be 
smaller than the EDM. 

In conclusion, we computed the deuteron MQM for various $\slashPT$ sources: 
the QCD $\tb$ term, quark EDM, quark and gluon chromo-EDMs,
and chiral-invariant four-quark operators. 
We performed these computations at leading order
in the framework of chiral EFT, with pions treated nonperturbatively.
The same parameters as in the corresponding calculation
of light-nuclear EDMs appeared here, 
except for the quark EDM, which involves an independent
short-range two-nucleon current.
While the results confirm the qualitative conclusions of Ref. \cite{Vri11b}, 
there are important quantitative differences. 
Due to its enhanced sensitivity to the QCD $\tb$ term, 
a potential measurement of the deuteron MQM would be 
complementary to one of the deuteron EDM. 
 
\noindent
{\bf Acknowledgments.}
We thank G. Onderwater for helpful discussions. 
UvK is grateful to KVI for hospitality.
This research was supported by the ROC NSC under grant NSC98-2112-M-259-004-MY3 (CPL), by the
Dutch Stichting FOM under programs 104 and 114 (JdV, RGET), and by the US DOE under contract DE-AC02-05CH11231 with
the Director, Office of Science, Office of High Energy Physics (EM),
and under grants DE-FG02-06ER41449 (EM) 
and DE-FG02-04ER41338 (EM, UvK).

\end{document}